\documentclass[aps,showpacs,showkeys,11pt,twocolumn]{revtex4}

\usepackage{graphicx,epic,eepic,epsfig,amsmath,latexsym,amssymb,verbatim,color}

\def\squareforqed{\hbox{\rlap{$\sqcap$}$\sqcup$}}
\def\qed{\ifmmode\squareforqed\else{\unskip\nobreak\hfil
\penalty50\hskip1em\null\nobreak\hfil\squareforqed
\parfillskip=0pt\finalhyphendemerits=0\endgraf}\fi}
\def\endenv{\ifmmode\;\else{\unskip\nobreak\hfil
\penalty50\hskip1em\null\nobreak\hfil\;
\parfillskip=0pt\finalhyphendemerits=0\endgraf}\fi}

\newcommand{\overt}{}
\newcommand{\bra}[1]{\langle{#1}|}
\newcommand{\ket}[1]{|{#1}\rangle}

\begin{document}
\title{Quantum error correction for continuously detected 
errors with any number of error channels per qubit}
\author{Charlene Ahn}
\email{cahn@theory.caltech.edu}
\affiliation{
Institute for Quantum Information,
Caltech 452-48,
Pasadena, CA 91125, USA}
\author{Howard Wiseman}
\email{H.Wiseman@griffith.edu.au}
\affiliation{Centre for Quantum Computer Technology,
Centre for Quantum Dynamics, School of Science,
Griffith University, Brisbane, QLD 4111, Australia}
\author{Kurt Jacobs}
\email{K.Jacobs@griffith.edu.au}
\affiliation{Centre for Quantum Computer Technology,
Centre for Quantum Dynamics, School of Science,
Griffith University, Brisbane, QLD 4111, Australia}

\begin{abstract}
It was shown by Ahn, Wiseman, and Milburn [PRA {\bf 67}, 052310 (2003)]
 that feedback control could be used as a quantum error correction
 process for errors induced by weak continuous measurement, given one
 perfectly measured error channel per qubit. Here we point out that this
 method can be easily extended to an arbitrary number of error channels
 per qubit.  We show that the feedback protocols generated by our method
 encode $n-2$ logical qubits in $n$ physical qubits, thus requiring just
 one more physical qubit than in the previous case.
\end{abstract}

\pacs{03.67.Pp, 42.50.Lc,03.65.Yz}
\keywords{quantum error correction, quantum feedback control}
\maketitle

Quantum error correction
\cite{shor-ec,steane-ec,knill-laflamme,gott-stab} and quantum feedback
control  \cite{wiseman_pra94,doherty-jacobs}
have a similar structure: a state of interest is measured, and then an
operation conditioned on the measurement is performed in order to
control the state. However, the end result of the control, as well as
the tools used to measure and control, are different for each. Quantum
feedback control has traditionally been used to control a known state
using weak measurements and Hamiltonian controls, whereas quantum
error correction uses projective measurements and unitary gates in
order to protect an unknown quantum state.

Despite the differences between the two techniques, the similarities
are sufficient to enable combining the two 
\cite{mabuchi-zoller,qec-spont,detected-jump1,ADL,detected-jump2,AWM}. 
In Ref. \cite{AWM}, we used feedback that was directly proportional to measured currents in
order to correct for a specific error process. In particular, we
assumed that the errors were \emph{detected}: the experimenter knows
precisely when and where errors have occurred because the environment
that produces those errors is being continuously monitored. Given this
assumption, we showed that feedback was able to protect a
stabilizer codespace perfectly for one perfectly measured channel per
physical qubit, and we discussed the results when the assumption of
perfect measurement is removed.

To be specific, in Ref. \cite{AWM} we analyzed the situation in which
given $n$ qubits, there was a single error channel on each qubit,
$E^{(j)}$ (on qubit $j$). That is, the decoherence of the register is
given by
\begin{equation}
\frac{d\rho}{dt } = \sum_j E^{(j) } \rho {E^{(j)}}^\dagger - \frac{1}{2 } \{ {E^{(j)}}^\dagger  E^{(j) } ,\rho   \} 
- i [H,\rho],
\end{equation}
where $H$ is an additional externally applied (``driving'') Hamiltonian.
Moreover, these errors could be perfectly detected in such a way that
the identity of the error (when and where it occurred) was known.  We
found that it was always possible to find feedback Hamiltonians and
``driving'' Hamiltonians that together perfectly corrected both the
error and the no-error evolution.  Our encoding used a single stabilizer
generator, i.e., encoding $n-1$ qubits in $n$.

In this paper we consider the following obvious generalization: What
happens if there are multiple channels $E^{(j),\alpha}$ on a single
qubit, all of which can  be detected?  (Here $j$ denotes the qubit
on which the channel acts, and $\alpha$ indexes which channel it is.)
Given a certain number of channels per qubit, what is the smallest
number of stabilizers needed to be able to use our protocol?
Equivalently, given $n$ physical qubits, how many logical qubits can be
encoded? To answer that question, we will first review the main result
of \cite{AWM}, in which there is only one channel per qubit (we will
drop the $\alpha$ index for clarity) and then present a generalization
of the results to the multiple-channel case.

In the detected-channel model, evolution is given by the error Kraus
operators 
\begin{equation}
\Omega_j = (E^{(j) } + \gamma^{(j)}) \sqrt{dt}
\end{equation}
and the no-error Kraus
operator 
\begin{eqnarray}
\label{eqn:omega0}
\Omega_0 &=& 1  - i H
   dt - \frac{1}{2 } {E^{(j)}}^\dagger E^{(j) } dt \nonumber \\
   &&  -\frac{\overt
   \gamma^{(j) } \overt}{2}(e^{-i \phi^{(j)}}
 E^{(j) } -   e^{i \phi^{(j)} } {E^{(j)}}^\dagger) dt ,
\end{eqnarray}
where $\gamma e^{-i \phi}$ is a complex parameter that describes the
kind of measurement (\emph{unraveling } of the master equation) that is
being done. That is, the {\em average } evolution reproduces the master equation
independently of $\gamma$:
\begin{equation}
\rho + d{\rho } = \Omega_0 \rho \Omega_0 ^\dagger + \sum_j \Omega_j \rho
\Omega_j^\dagger.
\end{equation}
For example,  $\gamma = 0$ for a Poisson unraveling, and $\gamma
\rightarrow \infty$ for a white-noise unraveling \cite{wiseman-semiclass}.  

In order to find
feedback Hamiltonians and driving Hamiltonians that together perfectly
correct both the error and no-error evolution, a sufficient condition
that needs to be met is
\begin{equation}
\label{eqn:kl}
\bra{\psi_i } D^{(j) } \ket{\psi_k } = 0,
\end{equation}
where $\ket{\psi_i}, \ket{\psi_k}$ are orthogonal states in the
codespace, and $D^{(j)}$ is the traceless part of $(E^{(j)} +
\gamma^{(j)})^\dagger (E^{(j)} + \gamma^{(j)})$.  Equation
(\ref{eqn:kl}) is just a variant of the Knill-Laflamme condition for correcting
errors \cite{knill-laflamme}, applied to the case in which the time and
position of the error are known. It is satisfied when the codespace
is generated by a stabilizer $S$ satisfying
\begin{equation}
\label{eqn:ec-gamma}
0 = \{S, D^{(j) } \}.
\end{equation}
 Since it is always
possible to find another Hermitian traceless one-qubit operator $s^{(j)}$
such that $\{ s^{(j)},  D^{(j) } \} = 0$, it then follows
that we may pick the single stabilizer generator
\begin{equation}
\label{eqn:oneS}
S = s^{(1) } \otimes \cdots \otimes s^{(n)}
\end{equation}
so that the stabilizer group is $\{1, S\}$. 

The identification of Eqn. (\ref{eqn:kl}) with the Knill-Laflamme
condition, combined with feedback results from \cite{wiseman_pra94}, show that
it is possible to correct the error using feedback.  Furthermore, the
no-error evolution given in (\ref{eqn:omega0}) can be corrected by
applying a driving Hamiltonian as follows:
\begin{equation}
\label{eqn:one-channel-H}
 H = \sum_j \frac{i}{2 } D^{(j) } S +\frac{i \overt \gamma^{(j) } \overt}{2}(e^{-i \phi^{(j)}}
 E^{(j) } -   e^{i \phi^{(j)} } {E^{(j)}}^\dagger).
\end{equation}
Putting this Hamiltonian in the total no-error Kraus operator in
(\ref{eqn:omega0}) gives, with $a=1 + O(dt)$, 
\begin{equation}
\Omega_0 = a 1 - \frac{1}{2 } \sum_j D^{(j) } (1 - S) dt.
\end{equation}
The second term here is zero on the codespace, so the no-error evolution does not
disturb the codespace.

For multiple channels  (denoted by $\alpha$)  on a given qubit, the
expressions in this previous work can easily be generalized. We are assuming
that the time scale of correction is fast compared to the time scale
of decoherence; therefore, different errors do not interfere with one
another, and all the expressions in our paper behave well (i.e.,
linearly). We should also note here that implicit in  the idea that
all errors are \emph{detected } is the assumption that, therefore,
given such a detection we know not only when and where ($j$) the error has
occurred, but also what the error is ($\alpha$).
In other words, given a
detection we can determine the error Kraus operator $\Omega_j^{\alpha}$  that has
been applied.

Given the above assumptions, to generalize to multiple-channel
protocols we must merely check whether for all $\alpha$ and $j$ it is
true that
\begin{equation}
\label{eqn:kl-gen}
\bra{\psi_i } D^{(j),\alpha } \ket{\psi_k } = 0.
\end{equation}
If (\ref{eqn:kl-gen}) holds, the corresponding errors
$E^{(j),\alpha}$ will be correctable, and we will see that this
condition also makes it possible to find a driving Hamiltonian such that
the no-jump errors are also corrected.

Let us first consider the case when there are two channels on a single
qubit: $\alpha = 1,2$. When there are two channels, a Bloch-sphere
analysis shows that it is possible to find a single $S$ such that
$\{S,D^{(j),\alpha}\} = 0$.  Let us consider qubit $1$: since
$D^{(1),1}$ and $D^{(1),2}$ are traceless, they can be represented by
two vectors on the Bloch sphere. In fact, $D^{(1),1}$ and $D^{(1),2}$
define a plane intersecting the Bloch sphere; now we pick $s^{(1)}$ to
be the operator that corresponds to the vector on the Bloch sphere that
is orthogonal to that plane. Since it is possible to find a unitary
rotation that takes $s^{(1)}$ to $\sigma_Z$ as well as $D^{(1),1}$ and
$D^{(1),2}$ to linear combinations of $\sigma_X$ and $\sigma_Y$, this
operator must anticommute with $D^{(1),1}$ and $D^{(1),2}$.  Doing the
same for the other physical qubits, we pick the single stabilizer
generator
\begin{equation}
S = s^{(1) } \otimes \cdots \otimes s^{(n)}
\end{equation}
so that the stabilizer group is $\{1, S\}$ as before. Again, this
procedure encodes $n-1$ qubits in $n$.

The next step is to consider three channels. Unfortunately, for three
 channels on a single qubit, it is not in general possible to
find a single $s$ which anticommutes with all the $D$ operators of the
channels; this is reflected by the fact that the Bloch sphere is
three-dimensional, and so given three arbitrary vectors, it is not
possible in general to find a fourth vector perpendicular to all three.

However, we can do almost as well. Let us return to (\ref{eqn:kl})
again. In fact for (\ref{eqn:kl}) to be true, it suffices to decompose
any given error operator, $D$, as $D = \vec{d } \cdot \vec{\sigma}$ and
to require
\begin{equation}
\label{eqn:kl-expand}
\bra{\psi_i } d_l \sigma_l \ket{\psi_k } = 0 \; \forall \; l.
\end{equation}
If our stabilizers are the two stabilizers of the familiar four-qubit
code for the erasure channel \cite{erasure},
\begin{eqnarray}
S_1 &=& XXXX, \nonumber\\
S_2 &=& ZZZZ,
\end{eqnarray}
we can see that for any $l$ one of these two, call it $S_{j(l)}$,  will satisfy
\begin{equation}
\label{eqn:mult-cond}
\{S_{j(l)}, \sigma_l \} = 0
\end{equation}
no matter what $D$ is, and thus
(\ref{eqn:kl-expand}) holds.

In this case,  with $a=1+O(dt)$ as before, we have 
\begin{equation}
\Omega_0 =  a 1 - \frac{D}{2} dt
 -\frac{\overt
   \gamma \overt}{2 } ( e^{-i \phi } E - e^{i \phi } E^\dagger) dt - i H dt.
\end{equation}
Let 
\begin{equation}
H = \sum_l \frac{i}{2} (d_l \sigma_l) S_{j(l) } + \frac{i \gamma}{2 } (e^{-i \phi } E -
e^{i \phi } E^\dagger),
\end{equation}
where $S_{j(l)}$ is defined as in
(\ref{eqn:mult-cond}). Then
\begin{equation}
\Omega_0 = a 1 - \frac{1}{2} \sum_l d_l \sigma_l (1 - S_{j(l)}),
\end{equation}
which leaves  the codespace invariant. This analysis is true for each
additional error channel we introduce. Thus no matter how many error
channels there are, as long as we can detect all of them and know
which error has happened and where the error has happened, we can
correct for the error and the no-error evolution.  This code encodes
two logical qubits in four physical ones.

In fact, this reasoning applies for $n$ qubits, where $n$ is even,
given the two stabilizers
\begin{eqnarray}
\label{eqn:gen_erasure_stab}
S_1 &=& X^{\otimes n}\nonumber\\
S_2 &=& Z^{\otimes n}.
\end{eqnarray}
Using these stabilizers with the constant Hamiltonian found above, it
is possible to encode $n-2$ qubits in $n$. 

This protocol, of course, borrows heavily from the stabilizer
formalism of the quantum erasure code. Indeed, the quantum erasure
code can be generalized using the stabilizers in
(\ref{eqn:gen_erasure_stab}) in the same way, with the same scaling of
$n-2$ logical qubits in $n$ physical ones; as far as we know this
scaling has not been explicitly noted in the literature. On the other
hand, our protocol differs from the erasure code in that we have made
a different and more restrictive assumption about the error model; as
a result, we only need to perform local measurements instead of a
highly nonlocal stabilizer measurement. To elaborate, the quantum
erasure code makes the same assumption that the position and time of
the error are both known. In the protocol given here, we make the
further assumption that we know \emph{what error } has occurred in
that measuring the error tells us what error has occurred. This
information about the error comes precisely from the detection of the
local measurements performed by the environment.  As in \cite{AWM},
these results indicate that if dominant error processes can be
monitored, using that information can be the key to correcting them,
and that the overhead in encoding is minimal (just two physical qubits).

\begin{acknowledgments}
C.~A. would like to thank the Centre for Quantum Computation for its
hospitality and acknowledge the support of an Institute for
Quantum Information fellowship.
\end{acknowledgments}

\end{document}